\documentclass[11pt]{article}
\usepackage[T1]{fontenc}
\usepackage{bm,amsmath,amssymb,mathtools,graphicx}
\usepackage{textcomp,gensymb,units,upquote}
\usepackage[a4paper]{geometry}
\usepackage{epsfig}
\usepackage[parfill]{parskip}
\usepackage[matrix,arrow,color]{xy}
\usepackage[usenames]{color}
\usepackage{mathrsfs}
\usepackage[font={small}]{caption}
\usepackage{subcaption}
\usepackage[export]{adjustbox}
\usepackage{float}
\usepackage{wrapfig}
\usepackage{microtype}
\usepackage{slashed}
\usepackage{braket}
\usepackage[pdftex]{hyperref}
\usepackage{wrapfig}
\usepackage{setspace}
\usepackage{lscape}

\input{xy}
\xyoption{all}

\input{epsf}

\makeatletter

\catcode`@=11
\renewcommand{\section}
{\@startsection{section}{1}{0pt}{\medskipamount}{\medskipamount}{\large\bf}}
\makeatletter\renewcommand{\subsection}
{\@startsection{subsection}{2}{\z@}{-3.25ex plus -1ex minus -.2ex}
{1.5ex plus .2ex}{\it }}

\numberwithin{equation}{section}
\catcode`@=12

\newcommand{\ban}{\begin{eqnarray}}
\newcommand{\ean}{\end{eqnarray}}

\newcommand{\Tr}{{\rm Tr}}

\newcommand{\IK}{\mathbb{K}}

\newcommand{\cH}{{\cal H}}
\newcommand{\cA}{{\cal A}}
\newcommand{\cE}{{\cal E}}

\newcommand{\cF}{{\cal F}}

\newcommand{\cU}{{\cal U}}

\newcommand{\sfa}{{\mathsf{a}}}

\newcommand{\complex}{{\mathbb C}} 
\newcommand{\zed}{{\mathbb Z}} 
\newcommand{\real}{{\mathbb R}} 

\def\e{{\,\rm e}\,}

\def\ii{{\,{\rm i}\,}}
\def\dd{{\rm d}}

\newcommand{\rank}{\mathrm{rank}}

\def\beq{\begin{equation}}
\def\bee{\begin{equation}}
\def\eeq{\end{equation}}
\def\bea{\begin{eqnarray}}
\def\eea{\end{eqnarray}}
\def\bd{\begin{displaymath}}
\def\ed{\end{displaymath}}

\newcommand{\Cint}{\int\kern-10.5pt-\kern7pt}

\newcommand{\be}{\begin{equation}}
\newcommand{\ee}{\end{equation}}
\newcommand{\bal}{\begin{align}}
\newcommand{\eal}{\end{align}}

\newcommand\fverbit{\egroup\item[\fbox{\unhbox\pippobox}]}
\newbox\pippobox

{

\def\be{\begin{equation}}
\def\ee{\end{equation}}
\def\bea{\begin{eqnarray}}
\def\eea{\end{eqnarray}}

}

\begin{document}

\begin{titlepage}
\setcounter{page}{1}

\vskip 5cm

\begin{center}

\vspace*{3cm}

{\Large OBSERVERS, LOCAL MEASUREMENTS, AND TOPOLOGY}

\vspace{15mm}

{\large Michele Cirafici}
\\[6mm]
\noindent{\em Dipartimento di Matematica, Informatica e Geoscienze, \\ Universit\`a di Trieste, Via A. Valerio 12/1, I-34127,
 \\ Institute for Geometry and Physics \& INFN, Sezione di Trieste,  Trieste, Italy 
}\\ Email: \ {\tt michelecirafici@gmail.com}

\vspace{15mm}

\begin{abstract}
\noindent 
An observer which propagates in a spacetime with dynamical gravity has access to an algebra of observables which is expected to be a von Neumann algebra with a trace. The observer can use projections from the algebra to perform local measurements. By using this fact we construct a simplicial complex out of measurements recorded on the worldline of the observer and we propose that such a complex captures topological information about spacetime, even when this is fluctuating.

\end{abstract}

\vspace{15mm}

\today

\end{center}
\end{titlepage}


\tableofcontents


\vspace{0.5cm}

\section{Introduction}

The presence of an observer might be necessary to properly describe the algebra of observables in perturbative quantum gravity in a closed universe. For example a careful analysis of the gravitational constraints in the case of de Sitter spacetime \cite{Chandrasekaran:2022cip} shows that in the presence of an observer the algebra of observables is deformed into a type $\mathrm{II}_1$ von Neumann algebra. In particular such an algebra has a trace, so that density matrices and entropies can be defined. The algebra contains also projections so that measurements can be implemented by the Born rule.

It was argued in \cite{Witten:2023qsv,Witten:2023xze} that such a situation might be generic and not special to de Sitter. In the de Sitter case, the main ingredients of the construction are the absence of asymptotic boundaries, which require an observer to define observables relationally, and the Hartle-Hawking no-boundary state, which is thermal and maximizes the entropy. It is natural to assume that an observer could play the same role in any closed universe, at least as a general mechanism to define observables. The Hartle-Hawking state, whenever it can be defined, is constructed universally from geometrical data and can be naturally interpreted as a universal thermal state. If it can also be interpreted as a maximum entropy state, then more general entropies can be defined as relative to such a  state. This state can then be used to define a trace on the algebra of the observables, thought of as an algebra of operators acting on the Hilbert space obtained from semiclassical quantization of spacetime, which is then a von Neumann algebra of type $\mathrm{I}$ or $\mathrm{II}$. Physically this means that the observer can describe the surrounding reality via a density matrix. The presence of the observer imposes strict conditions on the gravitational algebras of observables, with profound consequences both in de Sitter and in broader cosmological settings 
\cite{Abdalla:2025gzn,Akers:2025ahe,Blommaert:2025bgd,Chen:2024rpx,Chen:2025tbh,Cirafici:2024ccs,Geng:2024dbl,Geng:2025bcb,Harlow:2025pvj,Jensen:2023yxy,Kudler-Flam:2023qfl,Kudler-Flam:2024psh,Maldacena:2024spf,Speranza:2025joj}.

Another motivation to equip spacetime with an observer is that the algebra of observables defined along the worldline of the observer is a better defined concept than the algebra of observables defined in a spacetime region. Indeed the latter is somewhat ambiguous if spacetime is fluctuating, unless we specify  an invariant way to identify the region. In perturbative quantum gravity the metric itself is a quantum field, therefore a quantum state does not describe a sharply defined geometry but geometric quantities are subject to quantum fluctuations. Of course one can still work order by order in perturbation theory if a particular region can be specified in a gauge invariant way, such as the exterior of a black hole horizon. But this is not straightforward for arbitrary regions and geometrical quantities are difficult to interpret in the presence of dynamical gravitons. 

On the other hand the concept of an observer following a worldline has an operatorial meaning since the observer can measure quantities along their path and organize them according to their clock. Once an observer is introduced, local observables can be defined relationally. Then the general framework of quantum reference frames determines how the structure of the algebra of observables is affected \cite{AliAhmad:2024wja,DeVuyst:2024khu,DeVuyst:2024fxc,Fewster:2024pur,Hoehn:2023ehz}, roughly speaking by implementing constraint quantization with respect to the symmetry shared by the system and the observer.

In this note we propose that the observer can obtain information about the topology of spacetime by doing local measurements along the worldline even when spacetime is fluctuating. In perturbation theory the geometry of spacetime fluctuates but its topology is fixed, at a macroscopic scale, and can be in principle determined by the observer. The observer can do so by defining an abstract simplicial complex out of the measurement data: measurement events correspond to vertices which are connected by edges or higher dimensional simplices if they are correlated, in a precise sense decided by the observer. Since we only deal with measurement data, there is no need to define open sets in a fluctuating geometry or discuss their intersections, as one does in the nerve construction in algebraic topology. We propose that the homology of this complex captures topological information about the spacetime; more precisely about the part of spacetime accessible to the observer.

There is a certain ambiguity in the definition of the complex, due to the finite resolution of the measurement device and the time threshold used to define correlations. Such ambiguities can be tamed by using persistent homology, by varying the measurement parameters to obtain filtered complexes.

This paper has two goals: first we propose that the ordinary nerve construction used to define homology can be replaced in terms of abstract measurements performed by the observer; secondly we use persistent homology to investigate the stability of the construction under changes of the measurement parameters. We make no attempts at modelling a realistic observer or to apply our findings to realistic cosmological models, even if in principle our own Universe could have a non-trivial topology, with observable consequences; see \cite{Godet:2026gqw,Kajuri:2026yym,Philcox:2025faf} for a sample of the current literature.

This note is organized as follows. In Sections \ref{observers} and \ref{measurements} we discuss some aspects of gravitational algebras and local measurements and adapt them to the problem at hand. In Section \ref{topology} we discuss the construction of a simplicial complex associated with the observer performing measurements and discuss in detail a couple of examples. Section \ref{persistence} explains how by passing from complexes to filtered complexes one can remove certain ambiguities. We conclude with a few comments. An appendix contains some technical details.

\section{Observers and their gravitational algebras} \label{observers}

In this Section we will discuss the incorporation of an external observer in a closed spacetime and review why it affects the algebra of observables. We will then consider the particular case of de Sitter and detail a small generalization of this construction to include interactions between the observer and spacetime.

\subsection{Generalities}

Consider an observer travelling along a timelike curve $\gamma (\tau)$, parametrized by proper time $\tau$ in a spacetime $M$. While in ordinary QFT one can construct an algebra of observables  $\mathcal{A} (\cU)$ assigned to a spacetime region $\mathcal{U}$, this construction is problematic in perturbative quantum gravity. For example it is not clear what one would mean by $\mathcal{U}$ since the metric is fluctuating due to quantum effects; furthermore local observables are not diffeomorphism invariants but have to be gravitationally dressed. 

In the case of ordinary QFT in curved spacetime, whilst the observer propagates in spacetime, each value of the time parameter defines a region causally accessible to the observer. One expects that the observer can, for example, fix a time interval determined via a clock and measure observables in the associated causal region. However a more precise prescription involves the measurement of quantum fields along the observer's worldline. The algebra of observables along the worldline is generated by bounded functions of local quantum fields, appropriately smeared in the time direction, in such a way that OPE singularities are integrable \cite{Witten:2023qsv}. 

Without gravity the timelike tube theorem \cite{Strohmaier:2023hhy} assures that the algebra of operators $\mathcal{A} (\gamma)$ measured along the curve $\gamma$ agrees with the algebra of observables supported on a larger open set, the timelike envelope $\mathrm{Env} (\gamma)$. For example, if we fix two endpoints of the curve corresponding to a certain time interval, the corresponding $\mathrm{Env} (\gamma)$ is obtained by deforming the timelike curve $\gamma$ through a family of timelike curves while keeping the endpoint fixed. Under reasonable assumptions this algebra contains both local and extended operators. The algebra $\mathcal{A} (\gamma)$ defined along the worldline is an equivalent description for the algebras of observables defined on an open set.

In the presence of perturbative quantum gravity it is no longer unambiguous to talk about the algebra of observables supported on a spacetime region, unless this region is defined in an invariant way (say in the presence of a horizon). Nevertheless it is still meaningful to talk about the algebra of observables $\mathcal{A} (\gamma)$ measured along the observer's worldline, and indeed it was proposed in \cite{Witten:2023qsv,Witten:2023xze} that the latter is a more meaningful concept. Indeed it was argued in  \cite{Witten:2023xze} that the OPE algebra of operators defined along the observer worldline is a background-independent algebra. Once a spacetime is chosen, such an algebra can be represented as a Hilbert space algebra. The corresponding Hilbert space includes the quantum fields as well as the fluctuations of the metric, treated perturbatively.

The background-independent algebra includes the quantum fields $\phi (\gamma (\tau)) $ along the observer trajectory. The observer is described by some Hamiltonian $H_{\mathrm{obs}}$. If $H_{0}$ denotes the generator of a bulk diffeomorphism which fixes $\gamma$ but shift $\tau$ by a reparametrization, then the gravity constraint is
\be
\widehat{H} = H_{0} + H_{\mathrm{obs}} = 0 \, .
\ee
While the ``time'' evolution generated by the full Hamiltonian $\widehat{H} $ is a gauge symmetry (it would measure time from the point of view of someone external to the universe and it is therefore physically inaccessible), the bulk and the observer 
evolve in time relative to each other. Operators which commute with this constraint have the form
\be
\hat{\phi}_s = \Pi  \, \phi (\gamma (p + s)) \, \Pi \, ,
\ee
where the projector $\Pi$ ensures that the observer energy is bounded from below. Here $s$ is a constant and $p$ is the variable canonically conjugate with the observer's energy $q$, and it is physically interpreted as the observer's clock. Relational observables defined in this way are gauge-invariant. Such observables, together with the observer Hamiltonian (or better the observer's clock) form a background-independent algebra $\mathsf{A}_{\mathrm{obs}}$. This algebra has to be intended as an operator product algebra, characterized by its short distance singularities \cite{Witten:2023xze}. 

Once a spacetime $M$ is chosen, such algebra can be interpreted as an algebra of operators acting on a Hilbert space. Such a Hilbert space will in general include also metric fluctuations treated semiclassically in perturbation theory. Schematically $g_{\mu \nu} \sim g_{\mu \nu}^{(0)} + \sqrt{G} \delta g_{\mu\nu} $ and the metric becomes rigid in the strict $G = 0$ limit. In this case the algebra of observables can be completed to a Hilbert space algebra $\cA$ in the weak topology. The resulting von Neumann algebra is expected, and in some cases explicitly proven, to be a tracial algebra.

A possible reason for this is as follows \cite{Witten:2023xze}. If it can be defined, it is reasonable to expect that there is a distinguished state of the observer algebra: the Hartle-Hawking no boundary state $\Psi_{\mathrm{HH}}$ suitably generalized to include the observer. This state is defined formally by summing and integrating over all spacetime geometries and field theory data, with the prescription that each spacetime geometry has only one boundary on which the quantum state is defined. Since this prescription seems to be universal, it is reasonable to expect that such a state could be defined for every closed spacetime. If this state is also a state of maximal entropy, as it happens for example in de Sitter, then physical entropies can be defined relatively to such state. A state of maximal entropy, whose density matrix is the identity, can be used to define a trace as 
\be
\Tr \, \sfa = \braket{\Psi_{\mathrm{HH}} | \sfa | \Psi_\mathrm{HH}}
\ee
for every $\sfa \in \cA$. The cyclicity of the trace is then a consequence of the KMS-property of the Hartle-Hawking state  \cite{Witten:2023xze,Witten:2023qsv}. If a trace can be defined, then the algebra $\cA$ is necessarily of type $\mathrm{I}$ or $\mathrm{II}$. We expect it to be of type $\mathrm{I}$ if the observer has access to a complete Cauchy hypersurface, type $\mathrm{II}$ otherwise. Similarly if the trace state is not normalizable, and $\Psi_{\mathrm{HH}}$ is not really a state but a weight, the algebras are of type $\mathrm{I}_{\infty}$ or $\mathrm{II}_{\infty}$. In this note we will not need the full details of this construction.

While this might not be true in general, one would assume that in a reasonable spacetime (for which for example topological data can be defined) a causal observer, if it exists, is always capable of describing the environment with a density matrix and of performing measurements. We will therefore assume that our observer has projections and traces. Conservatively our analysis is limited to spacetimes for which this is true.

\subsection{More on de Sitter}

In our construction the observer is capable of measuring external fields along the worldline. To model precisely the measurement process is no longer a good approximation to assume the observer and the external fields are  non-interacting. However we will argue that for our purposes, studying coarse topological information about the surrounding spacetime, it is not necessary to introduce a detailed model of the observer-environment interaction. We will consider an interaction localized on the observer worldline and verify such localized interactions can be incorporated perturbatively in the relational observables. We will however not attempt to model a realistic detector.
 The purpose of this discussion is only to show that given a reasonably well understood modular setup for the observer and its environment which neglects the interactions between the two, including said interactions in perturbation theory does not alter significantly the construction. The specific results of this Section are not needed in the rest of the paper and the reader interested only in the relation between the observer and spacetime topology can skip this example.

Consider the Hamiltonian
\be
\widehat H_\lambda
=
\widehat H_0+\lambda V = H_{0}+q + \lambda V
\ee
where $V$ is an Hermitian interaction term, $V^\dagger = V$. $H_{0}$ is the bulk Hamiltonian and $q$ the observer's, with $[ H_0 , q] = 0$. The observer's Hamiltonian acts on an auxiliary Hilbert space $L^2 (\real)$ and its conjugate momentum $p$, so that $[q , p] = \ii$, represents the observer's clock. The observer action has the form $\int_\gamma \dd \tau \left( p \frac{\dd}{\dd \tau} q - \sqrt{g_{\tau \tau}} q \right)$, where $\tau$ is the observer proper time which parametrizes the worldline $\gamma$.

In the Hamiltonian formalism the constraint is represented on a reference time-slice, which we conventionally set at $\tau = 0$. We assume that the interaction term consists in a bulk field evaluated along the observer worldline, such as
\be \label{DeltaI}
\Delta I_{\mathrm{obs}}
=
-\lambda\int_\gamma d\tau\,\phi(\tau) \, .
\ee
In the interaction representation, the interaction at proper time $\sigma$ is therefore governed by the free Hamiltonian
\be
V(\sigma)
=
e^{i\widehat H_0\sigma}V e^{-i\widehat H_0\sigma}.
\ee
In the above example \eqref{DeltaI}, $V$ is just $\phi (\tau)$ evaluated at $\tau = 0$. In general it will be a complicated function of $\phi$ evaluated at the time slice $\tau = 0$. Note however that in general it needs not commute with $\widehat{H}_0$ (even if it commutes with $q$), and therefore will not represent an observable. As it is written $V$ is not a Hilbert space operator; we assume it to be suitably smeared and bounded to enjoy the necessary analyticity properties.

We want to impose the Hamiltonian constraint perturbatively in $\lambda$. For the time being let us forget about the positive energy projections $\Pi = \Theta (q)$. We now want to define a deformed relational observable
\be
\Phi_s 
=
\Phi_s^{(0)}
+
\lambda\Phi_s^{(1)}
+
O(\lambda^2)
\ee
defined order by order, with $\Phi_s^{(0)} = \phi (p+s)$, and such that 
\be
[\widehat H_\lambda,\Phi_s ]=0 \, .
\ee
At zeroth order in $\lambda$ gauge invariant observables have to commute with $\widehat{H}_0$. Since $H_{0}$ generates time evolution, $[H_0 , \phi(\tau)] = - \ii \frac{\dd}{\dd \tau} \phi (\tau)$, this can be compensated only when
\be
[q , \phi (\tau)] = [\ii \frac{\dd}{\dd p} , \phi (\tau)] = \ii \frac{\dd}{\dd \tau} \phi (\tau)
\ee
which is satisfied by setting $\tau = p + s$, for $s$ an arbitrary constant \cite{Witten:2023xze}. Therefore $\Phi_s^{(0)} = \phi (p+s)$. In other words, while $H_0$ generates the time evolution $\tau \rightarrow \tau + \epsilon$, $q$ shifts the observer clock $p \rightarrow p - \epsilon$.

At first order in $\lambda$ this becomes
\begin{equation} \label{1ordconstr}
[\widehat H_0,\Phi_s^{(1)}]
=
-[V,\Phi_s^{(0)}] \, .
\end{equation}
A similar problem arises when one incorporates gravitational corrections to the perturbative coupling of two von Neumann algebras associated one with an eternal black hole in AdS, and the other with an external bath, as was done in \cite{Cirafici:2024jdw}. Borrowing results from that article, the answer can be find using standard perturbation theory in the form
\be \label{1ordphi}
\Phi_s^{(1)}
=
-i
\int_{-\infty}^{0}
d\sigma\, \e^{\varepsilon \sigma}
e^{i\widehat H_0\sigma}
[V,\Phi_s^{(0)}]
e^{-i\widehat H_0\sigma}.
\ee
As before, $\tau = 0$ is the spatial slice where the Hamiltonian constraint is conventionally represented while $\Phi^{(1)}_s$ is a function of $p+s$ via $\Phi_s^{(0)} = \phi (p+s)$. We assume that the interaction $V$ has an appropriate profile for which this integral makes sense, for example is turned on adiabatically from the far past. Let us now check this expression directly. 

It is useful to rewrite \eqref{1ordconstr} as an ``evolution'' equation
\be
\delta (\Phi_s^{(1)}) = - \mathcal{J}_s 
\ee
where $\delta$ is the derivation $\delta = [ \widehat{H}_0 , \, \, \, ]$ and the source term is $ \mathcal{J}_s = [V,\Phi_s^{(0)}] $. To solve this equation one has to invert the operator $\delta$. However since $\delta$ is a commutator, it will have in general a non-trivial kernel. To cure this problem we use a $\ii \varepsilon$ prescription and deform this equation to 
\be  \label{1ordconstrdef}
[\widehat H_0,\Phi_s^{(1)}] - \ii \varepsilon \Phi_s^{(1)}
=
- \mathcal{J}_s\, .
\ee
with $\varepsilon$ positive. Now we compute the commutator $[\widehat H_0,\Phi_s^{(1)}]$ using the ansatz \eqref{1ordphi}
\begin{align} \label{check}
[\widehat H_0,\Phi_s^{(1)}] &= -i
\int_{-\infty}^{0}
d\sigma\, \e^{\varepsilon \sigma}
\left[ \widehat H_0 , 
e^{i\widehat H_0\sigma}
[V,\Phi_s^{(0)}]
e^{-i\widehat H_0\sigma} \right] \\ \label{check1}
& = -
\int_{-\infty}^{0}
d\sigma\, \e^{\varepsilon \sigma} \frac{\dd}{\dd \sigma} \left(
e^{i\widehat H_0\sigma}
[V,\Phi_s^{(0)}]
e^{-i\widehat H_0\sigma} \right)
 \\ \label{check2}
& = - \left. \left(  \e^{\varepsilon \sigma}
e^{i\widehat H_0\sigma}
[V,\Phi_s^{(0)}]
e^{-i\widehat H_0\sigma} \right) \right\vert_{-\infty}^0 + \varepsilon \int_{-\infty}^{0}
d\sigma\, \e^{\varepsilon \sigma}
\left(
e^{i\widehat H_0\sigma}
[V,\Phi_s^{(0)}]
e^{-i\widehat H_0\sigma} \right) 
\end{align}
where in \eqref{check2} we have integrated by parts and in \eqref{check1} we used the identity
\be
\frac{d}{d\sigma}
\left(
e^{i\widehat H_0\sigma}
A
e^{-i\widehat H_0\sigma}
\right)
=
i
e^{i\widehat H_0\sigma}
[\widehat H_0,A]
e^{-i\widehat H_0\sigma} 
= 
i [\widehat H_0,
e^{i\widehat H_0\sigma}
A
e^{-i\widehat H_0\sigma}]
\ee
valid for any operator $A$ which does not depend on $\sigma$. Now we evaluate
\be
\left. \left(  \e^{\varepsilon \sigma}
e^{i\widehat H_0\sigma}
[V,\Phi_s^{(0)}]
e^{-i\widehat H_0\sigma} \right) \right\vert_{-\infty}^0=  [V,\Phi_s^{(0)}] - \lim_{\sigma \rightarrow - \infty} \left(  \e^{\varepsilon \sigma}
e^{i\widehat H_0\sigma}
[V,\Phi_s^{(0)}]
e^{-i\widehat H_0\sigma} \right) 
\ee
where the last term vanishes due to the damping factor. Finally, if we note that the last term of \eqref{check2} is precisely $\ii \varepsilon \Phi_s^{(1)}$, we have shown that the ansatz \eqref{1ordphi} indeed satisfy \eqref{1ordconstrdef}. Assuming the adiabatic limit exists, we can now remove $\varepsilon$.

Putting everything together
\be
\Phi_s
=
\phi (p+s)
-
i\lambda
\int_{-\infty}^{0}
d\sigma\,
e^{i\widehat H_0\sigma}
[V,\phi (p+s)]
e^{-i\widehat H_0\sigma}
+
O(\lambda^2) \, ,
\ee
obeys the Hamiltonian constraint at first order in $\lambda$. It is clear how the definition of this relational observable can be extended to all orders in perturbation theory in $\lambda$, in the form of a Dyson series. The full perturbative extension is discussed in Appendix \ref{appendix}, where we show that the above construction of the operator algebra associated to the observer survives the inclusion of interactions order by order.

Finally to obtain expressions which make sense in the type $\mathrm{II}_1$ algebra, we have to multiply by the projection $\Pi = \Theta (q)$ to obtain $\Pi \Phi_s \Pi$. This is straightforward as long as the interaction term commutes with $\Pi$, that is it preserves the condition that the observer's Hamiltonian is bounded.

We will now argue that the new algebra $\cA_\lambda$ is again a type $\mathrm{II}_1$ von Neumann algebra. Denote by $\cA_0$ the type $\mathrm{II}$ algebra obtained with $\lambda = 0$ \cite{Chandrasekaran:2022cip}. The algebra $\cA_0$ has a trace and a maximal entropy state $\Psi_{\mathrm{max}} = \Psi_{\mathrm{dS}} \otimes \e^{-\beta_{\mathrm{dS}} q / 2} \sqrt{\beta_{\mathrm{dS}}}$. Araki's theory of perturbation of KMS states guarantees that under determinate analytical conditions, which we assume are met and essentially amount to the perturbation being small enough, after the perturbation the theory settles into another KMS state \cite{bratteli}. Note that this implicitly assumes that we are really dealing with some bounded function of $\phi$, not the quantum field itself

Indeed consider the deformed state
\be \label{VdefKMSpsi}
\psi_V = Z_V^{-\frac12}  \e^{-\beta_{\mathrm{dS}} (H_0 + \lambda V)/2} \Psi_{\mathrm{dS}} \, ,
\ee
with $Z_V = \Vert  \e^{-\beta_{\mathrm{dS}} (H_0 + \lambda V)/2} \Psi_{\mathrm{dS}}  \Vert^2$, and the corresponding state for the bulk-observer system
\be \label{VdefKMS}
\Psi_{V} = Z_V^{-\frac12} \e^{-\beta_{\mathrm{dS}} (H_0 + \lambda V)/2} \Psi_{\mathrm{max}} \, .
\ee
Denote with $\Delta = \e^{-\beta_{\mathrm{dS}} H_0}$ and $J$ the modular operator and the modular conjugation of the unperturbed bulk algebra $\cA_{\mathrm{bulk}}$ and recall that $\Psi_{\mathrm{dS}}$ is its modular vector. Note that in \eqref{VdefKMS} the observer is just a factor.

We can check that $\psi_{V}$ is fixed by the modular operator $\Delta_V = \e^{- \beta_{\mathrm{dS}} L_V}$, $\Delta_V \psi_{V} = \psi_{V}$, where the Liouvillean operator is $L_V = H_0 + \lambda V - \lambda J V J$, with $J$ the modular conjugation. Note that $H_0 + \lambda V$ is the perturbed generator of the dynamics. Both $H_0 + \lambda V$ and $L_V$ generate the same automorphism group on the bulk algebra
\be
\e^{\ii t (H_0 + \lambda V)} \sfa \e^{- \ii t (H_0 + \lambda V)} = \e^{\ii t L_V} \sfa \e^{- \ii t L_V} \, , \qquad \sfa \in \cA_{\mathrm{bulk}} \, .
\ee
To begin with we can check that  
\begin{align}
 \e^{\ii (H_0 + \lambda V) t} \left(H_0 + \lambda V - \lambda JVJ \right) \e^{-\ii (H_0 + \lambda V) t} \Psi_{\mathrm{dS}}
& = \left[ H_0 + \lambda V  - \e^{\ii H_0  t} \left( \lambda JVJ \right) \e^{-\ii H_0  t} \right] \Psi_{\mathrm{dS}}
\nonumber \\  & =\lambda \left( V - \e^{\ii H_0  t} J V \right) \Psi_{\mathrm{dS}}
\end{align}
where we have used that the state $\Psi_{\mathrm{dS}}$ is annihilated by $H_0$ and that $J V J$ belongs to the commutant algebra of $\cA_{\mathrm{bulk}}$\footnote{ 
Indeed let $B$ an element of the commutant algebra. In particular $B (t) = \e^{i t H_0} B \e^{- \ii t H_0}$ since the unperturbed evolution preserve the algebra and its commutant. But since
\[
\frac{\dd}{\dd t} B (t) = \ii [H_0 , B(t)] =  \ii [H_0 +\lambda V , B(t)] \, ,
\]
it follows that $\e^{\ii t \left(H_0 +\lambda V \right)} J V J \e^{-\ii t \left( H_0 +\lambda V \right)}=\e^{\ii t H_0} J V J \e^{-\ii t H_0} $ since both are solutions of the same differential equation with the same initial condition.
}.

Now we use the KMS assumption to analytically continue $t = - \ii \beta_{\mathrm{dS}} /2$. Recall that $\e^{\ii t H_0 }  = \Delta^{- \ii t/\beta_{\mathrm{dS}}}$ in terms of the unperturbed modular operator. Therefore
\begin{align}
\left( V - \Delta^{-1/2} J V \right)  \Psi_{\mathrm{dS}}=  \left( V - J \Delta^{1/2} V \right)  \Psi_{\mathrm{dS}} = \left( V - V^\dagger \right) \Psi_{\mathrm{dS}} = 0
\end{align}
since by assumption the perturbation is Hermitian $V = V^\dagger \in \cA_{\mathrm{bulk}}$. In the computation we have freely used the following properties of the bluk modular operator and modular conjugation: $J^2 = 1$, $J \Delta^{1/2} J = \Delta^{-1/2}$ and $J \Delta^{1/2} A \Omega = A^\dagger \Omega$ for any operator $A$ and modular vector $\Omega$. 

Putting all together, now we see that
\be
 \left( H_0+ \lambda V - \lambda JVJ \right) \e^{-\ii (H_0 + \lambda V) (-\ii \beta_{\mathrm{dS}}/2)} \Psi_{\mathrm{dS}}  = 0 \, ,
\ee
which means that \eqref{VdefKMSpsi} is annihilated by $L_V =  H_0 + \lambda V - \lambda JVJ$. Since it is by assumption a KMS state, it follows that the bulk vector $\psi_V$ is also cyclic and separating with modular operator $\Delta_V = \e^{- \beta_{\mathrm{dS}} L_V }$. 

If we now consider the vector \eqref{VdefKMS}, we can see easily that also $L_V \Psi_V = 0$ since the observer state is just a factor. This fact is needed to define a trace. See \cite{Cirafici:2024jdw,Cirafici:2024itu} for a similar computation. 

Note that from the explicit expression of \eqref{VdefKMS}, it is the size of the perturbation $|| \lambda V ||$, assumed finite, which settles the distance in state space between the perturbed and the unperturbed state $|| \Psi_V - \Psi_{\mathrm{max}} ||$. Physically we would expect $\Psi_{V}$ to have ``lower entropy'' than $\Psi_{\mathrm{max}}$ since interactions will induce correlations between the observer and the bulk and the resulting state will not be maximally mixed. However that this fact is not captured by the present formalism; we cannot really compare the entropies between the two algebras since in type $\mathrm{II}$ algebras they are only defined up to an additive constant.

Now we can define the algebra by imposing the physical constraint $\widehat{H}_\lambda = 0$
\be
\left( \cA_{\mathrm{bulk}} \otimes B (L^2 (\real)) \right)^{\widehat{H}_\lambda} \, .
\ee
Note that since $\widehat{H}_\lambda$ and $L_V + q$ generate the same automorphism, one can also impose the constraint by using $L_V + q$. This is useful since as we have just shown $L_V \Psi_V = 0$ and one can follow step by step the procedure discussed in \cite{Chandrasekaran:2022cip,Witten:2023qsv,Witten:2023xze}; among other things one imposes a projection $\Pi = \Theta (q)$ which ensures that the observer energy is positive
\be
\cA_\lambda = \Pi \, \left( \cA_{\mathrm{bulk}} \otimes B (L^2 (\real)) \right)^{\widehat{H}_\lambda} \, \Pi \, .
\ee
The algebra $\cA_\lambda$ is of type $\mathrm{II}_1$. In particular the vector $\Psi_V$ can be used to define a deformed trace on the algebra $\cA_{\lambda}$
\be
\Tr_\lambda \sfa_V = \braket{\Psi_V \vert \sfa_V \vert \Psi_V}
\ee
for any operator $\sfa_V \in \cA_\lambda$. The fact that this is indeed a trace follows from the crossed-product construction; in particular the cyclicity of the trace follows from the KMS property of the normalized perturbed state $\psi_V$, from a direct application of an argument of \cite{Witten:2023xze}.

In summary one can check in the case of de Sitter that introducing a coupling between the observer and the bulk does not change qualitatively the structure of the algebra of observables. Similarly, whenever the general structure of this algebra can be generalized to arbitrary spacetimes, for example following \cite{Witten:2023xze}, we expect that the reasonings outlined above are sufficiently general to hold.

Since the aims of this paper are less ambitious, in the following we will assume that interactions between the observer and the surrounding spacetime, especially those interactions occurring during measurements along the observer's worldline, are implicitly part of the formalism. We will however not explicitly include them in the notation to avoid clutter. In the next Sections we will use effects representing localized measurements accessible to the observer.

\section{Quantum measurements} \label{measurements}

We will now describe how concretely an observer can perform measurements when their algebra is equipped with a faithful trace. Consider a von Neumann algebra $\cA$ of type $\mathrm{II}$ or type $\mathrm{I}$, and assume the algebra has a faithful trace. States are described by complex-valued linear functionals $\omega \, : \, \cA \longrightarrow \complex$ which are positive and normalized, $\omega (1) = 1$. A state $\omega$ is positive if for all $\sfa \in \cA$ we have $\omega ( \sfa^\dagger \sfa) \ge 0$. Since the algebra $\cA$ has by assumption a trace, certain states can be represented by density operators
\be
\omega_\rho (\sfa) = \Tr \, \sfa \rho \, \qquad \sfa \in \cA
\ee
which are positive and normalized $\Tr \rho = 1$. Such states are called normal.

Among observables in $\cA$ are projection operators, corresponding to Yes/No measurements. A measurement with two outcomes can be set up via the projection operators $\Pi$ and $1 - \Pi$. A projection operator is a self-adjoint operator such that $\Pi^2 = \Pi = \Pi^\dagger$. Projection operators form a lattice $\mathsf{Proj} (\cA)$ under a partial ordering operation defined as follows: $\Pi_1 \preceq \Pi_2$ if $\Pi_2 - \Pi_1$ is a positive operator. Equivalently $\Pi_1 \cH \subseteq \Pi_2 \cH$. 

The main difference between type $\mathrm{II}$ and type $\mathrm{I}$ factors is the absence of minimal projections. In type $\mathrm{II}$ factors every nonzero projection can be further decomposed into smaller nonzero projections. In particular such algebras do not admit normal pure states. The size of a projection can be quantified via its trace; in a type $\mathrm{II}_1$ factor one has
$
0\le \Tr (\Pi)\le 1,
$
and every value in this interval can occur. 

The spectral theorem states that given a bounded self-adjoint operator $\sfa$ in a von Neumann algebra $\cA$, there exists a unique projection-valued measure $E_\sfa : B (\real) \longrightarrow \mathsf{Proj} (\cA)$, where $B (\real)$ denotes the Borel subsets of $\real$, such that
\be
\sfa = \int_{\real}  \lambda \,  \dd E_\sfa (\lambda)
\ee
in the strong operator topology. More generally if the observable has spectrum $\sigma (\sfa)$ the projection-valued measure is defined on $B (\sigma (\sfa))$. 

Assume now that $g : \real \longrightarrow \complex$ is a bounded Borel function. Then functional calculus allows to construct
\be
g (\sfa ) = \int_{\real} g (\lambda) \dd E_\sfa (\lambda)
\ee
as an operator in the von Neumann algebra $\cA$. In particular even if the operator $\sfa$ is not bounded we can choose the function $g$ in such a way that $g (\sfa)$ is bounded. From now on we will assume that all our operators are bounded or replaced by bounded functions, even if this is not apparent from the notation. Note that the bicommutant theorem of von Neumann algebras implies that  if $\sfa \in \cA$, then also all its spectral projections and bounded functional calculus also belong to $\cA$, since they commute with all the operators which commute with $\sfa$.

Consider any self-adjoint observable $\sfa \in \cA$ with projection valued measure $E_\sfa$ and spectrum $\sigma (\sfa)$. Assume that the system is in a certain state $\omega_\rho$. Then according to the Born rule the probability that a measurement of the observable $\sfa$ yields a value in a Borel set  $\Delta \in B (\sigma (\sfa))$ is given by
\be
p (\Delta) = \omega_\rho \left( E_\sfa (\Delta) \right) = \Tr \left( \rho E_\sfa (\Delta) \right) \, .
\ee

We can use the functional calculus to define a certain class of projections. Introduce the indicator function $\mathbf{1}_{[a , \infty )} : \real \longrightarrow \{ 0 , 1 \}$ so that $\mathbf{1}_{[a , \infty )} (\lambda)$ is $1$ if $\lambda \ge a$ and zero otherwise. Now by using functional calculus we can define
\be
\mathbf{1}_{[a , \infty )} (\sfa) = \int_{\real} \mathbf{1}_{[a , \infty )} (\lambda) \dd E_\sfa (\lambda) = E_\sfa ([a , \infty))
\ee 
which physically is the spectral projection of the operator $\sfa$ onto the part of its spectrum above a certain threshold $a$. Since this is a projection, its eigenvalues are only $0$ and $1$, corresponding to the physical statement that the result of a measurement of the observable $\sfa$ lies in the interval $[a , \infty)$ or not. In other words, if a system is in a state described by a density matrix $\rho$, then 
\be
\Tr \left( \rho \mathbf{1}_{[a , \infty )} (\sfa) \right) = \Tr \left( \rho \, E_\sfa ( [a , \infty )) \right)
\ee
is the probability that a measurement of the observable $\sfa$ yields an eigenvalue $\lambda \ge a$.

It is useful to keep in mind that the measurement prescription we have discussed so far can be generalized.  In generalized measurements, the fundamental objects describing outcomes are so-called \textit{effects}. An effect is an operator $\mathcal{F} \in \cA$ such that $0 \le \cF \le \mathbf 1$. The quantity $\Tr \rho \, \cF$ is still interpreted as the probability that the corresponding event occurs. A set $\{ \cF_a\}$ is called a \emph{POVM} (positive operator valued measure) if
\be
\cF_a \ge 0, \qquad \sum_a \cF_a = \mathbf 1 .
\ee
Each $\cF_a$ corresponds to one possible measurement outcome. Ordinary measurements are recovered in the case where all effects are projections. POVMs describe unsharp measurements arising from realistic detectors with finite efficiency and noise. Physically, an effect can be derived from a microscopic detector model by coupling a detector to the field and tracing out internal detector degrees of freedom. 

From the previous discussion we can construct a simple example of an effect by replacing the indicator function with a smooth function. A possibility is a logistic function
\be
f_{a,\varepsilon}(\lambda)=\frac{1}{1+\e^{-(\lambda-a)/\varepsilon}} \, ,
\ee
where $\varepsilon > 0$ corresponds to a detector with finite resolution. The effect is then defined via  functional calculus as $\cF = f_{a , \varepsilon} (\sfa)$. A simple POVM is then the two-outcome set $\{ \cF , 1 - \cF \}$.

Note that after a measurement within a type $\mathrm{II}$ algebra, the collapsed state cannot be pure, since the projection can always be further decomposed. This is in stark contrast with type $\mathrm{I}$ algebras. Of course the physical interpretation of a measurement is independent of this and the collapse is still not invertible.

\section{Reconstructing topological information} \label{topology}

Heuristically we expect that while travelling through a spacetime region, an observer can gain information about the surroundings via measurement devices. Measurements recorded along the worldline correspond to physical events happening within that region. Realistic observers could be for example a laboratory travelling alongside the Earth or a cosmonaut recording observations within their spacecraft, moving along a trajectory which is approximately a timelike geodesic. The observer can then detect events, such as the emission of elementary particles by a nearby star, by a repeated series of measurements, and from these reconstruct information about their environment.

We would like however discuss an idealized situation where perturbative quantum gravity effects are not-negligible, even if small. In such a situation we cannot in general talk about a spacetime region since geometry is fluctuating. However, as discussed before, the algebra of observables $\cA (\gamma)$ measured along the observer worldline is a natural stand-in for the algebra of observables supported on a spacetime region, even when the latter concept is ambiguous.

In this Section we will describe how the observer can obtain topological information about spacetime by doing measurements of observables in $\cA (\gamma)$. Heuristically the idea is as follows. Operationally we want to replace a region of spacetime with an interval of time during which certain measurements take place. Even when spacetime regions become ambiguous due to quantum fluctuations, proper time intervals as measured by the observer remain well defined. Two such ``regions'' overlap if the corresponding intervals correspond to above-threshold probabilities for at least one common detector and are close enough in proper time; corresponding for example to repeated measurements of the same observable.  Similarly one can talk about triple or higher overlaps. By sharpening these ideas we will define a simplicial complex whose homology can be interpreted as providing topological information about spacetime.

This construction is inspired by the article \cite{curto}, where it is shown how combinatorial patterns in neural activity are associated with the external environment.

\subsection{Simplicial homology from local measurements}

Let us assume that the observer is travelling in a spacetime $M$ along a worldline $\gamma (\tau)$ parametrized by proper time $\tau$. Fixing a spacetime $M$ corresponds to choosing a Hilbert space representation $\cA$ of the background-independent algebra of observables. By taking closure in the weak operator topology we can assume that such an algebra is a von Neumann algebra. For reasons discussed in Section \ref{observers} we will assume that the algebra is of type $\mathrm{II}$. For the observer we assume the model outlined in Section \ref{observers}, eventually taking into account interactions in such a way that the observer can interact with the surrounding spacetime via idealized measurements. In general, when taking into account perturbative quantum gravity, such interactions will induce a backreaction, less negligible the more the measurement process is complicated. However in this note we are only interested in topological information and it is safe to assume that the measurement process will not induce topology changing transitions. We will therefore ignore backreaction in the following.

We assume that the observer is equipped with a measurement device. Such a device consists of several detectors which can be used to measure observables along the observer's worldline. Associated to the detectors there is a collection of effects $\{ \cE_i \}$ which we will call response operators. These operators are chosen by the observer according to their physical model of the accessible universe and correspond to the observables they want to measure. Each response operator has a ``yes'' or ``no'' outcome described by the two-ouctome POVM  $\{ \cE_i , 1 - \cE_i \}$. 

Measurements done by the observer will be referred to a state on the type $\mathrm{II}$ von Neumann algebra represented by a density operator $\rho \in \cA$. This state describe a global state of spacetime plus the observer, eventually reduced to the region of spacetime that the observer can causally communicate with. Eventually we can consider that an ``ancilla'' is part of this state and it is used by the observer for the measurement process. 

At proper time $\tau$ the observer computes the probability $p_i (\tau) $ that one of the response operators $\cE_i$ gives a ``yes'' outcome. We say that we have a detection event when this probability is above a certain threshold chosen by the observer
\be \label{Ethreshold}
p_i (\tau) = \mathrm{Tr} \left( \rho \, \cE_i (\tau) \right) = \omega_\rho (\cE_i (\tau)) > \theta_i \, ,
\ee
where the trace is determined by the maximum entropy state. Colloquially we say that the associated detector is \textit{active}. As the observer moves in spacetime the probability $p_i (\tau)$ will in general depend on proper time $\tau$ via the dependence on time of the algebra of observables. At time $\tau$, $p_i (\tau)$ is the probability that  the two-outcome measurement $ \{ \cE_i(\tau),1-\cE_i(\tau) \}$ gives the ``yes'' outcome.

Note that in a realistic setup the observer would record the detector outcomes and use them to infer the probabilities statistically. In our idealized construction we will work directly with the theoretical probabilities $p_i (\tau)$. 
What the observer really wants is to build a model of the external world. To do so they develops physical theories, which are here represented by the operators $\cE_i$, and compute the probabilities \eqref{Ethreshold}. 
The threshold $\theta$ can then be interpreted as the confidence level that a particular mathematical model actually corresponds to the click of the detector. It is therefore via the probabilities \eqref{Ethreshold} that the observer describes the surrounding spacetime. Equivalently one can take a frequentist perspective, where the observer repeats the measurement many times and rejects outcomes as statistical flukes if they happen with frequency below a certain threshold.

We assume that the observer can continue to compute the probabilities \eqref{Ethreshold} with the same state $\rho$, which is a very complicated state which describes the universe plus the observer. In principle after a measurement the state $\rho$ should be updated to $\Pi \rho \Pi$, up to a normalization, for example if performing a projective measurement characterized by a projection $\Pi$. However the density operator $\rho$, which describes a state of the Universe plus the observer, is naturally expected to be a state of very large entropy. For example in the case of de Sitter, such entropy is the generalized entropy, which includes the cosmological horizon area divided by $4 G$. While in general $\rho$ and $\Pi \rho \Pi$ can be very different states, their entropy only differs by the amount of information gained by the observer with their measurement. The difference between the pre- and post- measurement states involves only a tiny fraction of information about the observer algebra. Therefore we expect that only extremely complicated operators\footnote{One can make a more clean statement in the context of black holes in AdS/CFT. In that case, to engineer a black hole, one can start with a pure state and follow its gravitational collapse. In the large $N$ limit, semiclassical gravity gives the usual picture of a thermal state with a large entropy given by the area of the horizon. But due to unitarity, the state is still pure in the full fledged boundary CFT. However this is very difficult to see in practice, since any operator which could do so must have a parametrically large complexity which scales with $N$. This can be made quite precise by looking at the algebra of observables \cite{Chandrasekaran:2022eqq}}, of complexity of order of the exponential of the entropy, can resolve the difference between $\rho$ and $\Pi \rho \Pi$. A realistic assumptions is that the response operators which constitute the detector of the observer are on the other hand very simple (for example the polarization of a photon) and will therefore not see the difference between the pre- and post-measurement state. This is analogous to the usual semiclassical treatment of measurements in large-entropy systems, where simple observables cannot distinguish states differing by a small amount of microscopic information. We will therefore assume that the operators $\cE$ are sufficiently simple that we can continue to compute the probabilities \eqref{Ethreshold} with the same state $\rho$. While this argument is only heuristic, a more realistic construction where we take into account the post-measurement state is possible, but would complicate considerably the construction without altering the main conceptual point. 

Realistic detectors will have a finite temporal resolution. Therefore the measurement process will not be instantaneous but will take a certain amount of time, as registered by the observer's clock. We define a measurement event as an open interval of proper time $I_\alpha = (a_\alpha , b_\alpha) \subset \real_\tau$ for which at least one detector is active. During a measurement event $I_\alpha$ we denote by
\be
S_\alpha = \left\{ 
i \in \left\{ 1 , \dots , N \right\} \, \vert \, \exists \, \tau \in I_\alpha \, \text{such that} \, p_i (\tau) > \theta 
\right\}
\ee
the set of labels of active detectors, with $N$ the number of detectors. Here $\theta \in [0,1)$ is an arbitrary threshold, chosen common for each detector for simplicity. In plain words measurement events correspond to time intervals during which at least one detector has an above threshold ``yes'' probability. There is of course an ambiguity in choosing $\theta$, which we will address in the next Section.

We will assume for simplicity that the intervals $I_\alpha$ are not overlapping, although it is not essential. The intervals are finite in number and ordered according to their time. The label $\alpha$ takes values in an ordered set $J$. For simplicity we will assume that $J$ has finite cardinality, corresponding to an the observer who performs finitely many measurements. To a measurement event we associate a vertex. The collection of vertices is therefore given by
\be
\mathsf{V}_\theta  = \left\{ \alpha \in J \,  \vert \, S_\alpha \neq \emptyset \right\} \, .
\ee
For $\alpha \in \mathsf{V}_\theta$, we denote by $v_\alpha$ the corresponding abstract vertex.
Thus $\mathsf{V}_\theta$ is 
the set of labels corresponding to measurement events for which at least a detector has an above-threshold ``yes'' probability. By construction $\mathsf{V}_\theta$ is finite. Note that in physical situations the number of vertices will in general be larger than the number of detectors; the same detector can be active during two separate measurement events, leading to two distinct vertices. 

Consider now a collection of measurement events $\{ I_{\alpha_0} , \cdots , I_{\alpha_k} \}$, each of the form $I_{\alpha_j} = (a_{\alpha_j} , b_{\alpha_j}) $. We define its overall duration as
\begin{align}
\delta_{ \{ I_{\alpha_0} , \cdots , I_{\alpha_k} \} }  =\max_{0 \le j \le k} {b_{\alpha_j}} -  \min_{{0 \le j \le k}} {a_{\alpha_j}}  \, .
\end{align}
This is simply the difference between the proper time where the latest measurement ends and the time where the earliest one begins. This notation avoids the need to discuss an ordering of the vertex set. When measurement events correspond to vertices we will often use the equivalent notation $\delta_{  \{ v_{\alpha_0} , \cdots , v_{\alpha_k} \} }$, or $\delta_{\sigma}$ for $\sigma \subseteq \mathsf{V}_\theta$.

Two or more subsequent measurements can record the same physical event. The condition for this to be the case is that they share at least a common active detector, that is  
\be
\bigcap_{j=0}^k S_{\alpha_j} \neq \emptyset \, ,
\ee
during a prescribed period of time determined by a parameter $\Lambda$, which we assume larger than all the intervals $I_{\alpha}$. This period of time is chosen by the observer and might of course lead to spurious effects, to be discussed in the next Section.

Out of these data we define a simplicial complex as
\be
\mathsf{K}_{\theta , \Lambda} = \left\{
\sigma \subseteq \mathsf{V}_\theta \, \vert \sigma \ \text{is finite} \, , \bigcap_{\alpha \in \sigma} S_{\alpha} \neq \emptyset \, ,
\text{and} \,  \delta_{\sigma}  \le \Lambda
\right\} \, ,
\ee
By conventions we add the empty simplex to the definition of $\mathsf{K}_{\theta , \Lambda}$. This is the simplicial complex whose vertex set is $\mathsf{V}_\theta$ and where a set of vertices $\{ v_{\alpha_0} , \cdots , v_{\alpha_k} \}  $ defines a $k$-simplex if and only if $S_{\alpha_0} \cap \cdots \cap S_{\alpha_k} \neq \emptyset$ and the total duration  of the intervals does not exceed $\Lambda$. Note that this construction parallels the nerve complex of a family of open sets in algebraic topology. However, since we cannot work directly with spacetime regions when geometry is fluctuating, we have rephrased this construction in terms of measurements carried out by the observer.

An edge connects two vertices $v_\alpha$ and $v_\beta$ if $S_\alpha \cap S_\beta \neq \emptyset$ and $\delta_{ \{ v_\alpha , v_\beta \} } \le \Lambda$.  A simplex $\sigma \in \mathsf{K}_{\theta, \Lambda}$ is a collection of measurement events which share one or more common active detectors within the prescribed time window set by $\Lambda$.
We stress that $\Lambda$ is not a constraint on the number of measurements, which can be arbitrary large, but sets the timescale for which subsequent measurements can be considered correlated. It is only needed since a realistic observer has only a finite number of detectors and the same detector can be used to measure two different phenomena, for example separated in time. The parameter $\Lambda$ is chosen by the observer according to their mathematical models of reality. A simplex $\sigma \in \mathsf{K}_{\theta, \Lambda}$ should be thought of as a stand-in for intersecting open sets in algebraic topology when the latter concept is not well defined, due to geometry fluctuating. The simplicial complex does not live on spacetime but is a purely combinatorial object constructed in abstract terms from information associated with the observer worldline $\gamma$.

The observer has the freedom to choose the parameters $\theta$ and $\Lambda$. In doing so the observer decides the thresholds for which they can declare that an event has happened (the probability of a ``yes'' outcome is above the threshold) and that two measurements are closed enough in time to be declared describing the same physical event. This leaves some room for ambiguities which we will address later by varying these parameters. 

Let us now check that this simplicial complex is well defined. It follows from the definition that the complex does not depend on the order of the $v_{\alpha_j}$, and neither does the temporal span $\delta_{ \{ I_{\alpha_1} , \cdots , I_{\alpha_k} \} }$. The fact that the measurement events are ordered in time is immaterial; the observer notes them down and only afterwards uses them to construct the simplices. 

More importantly every face of a simplex is again a simplex. Indeed consider any subfamily $\{ I_{\alpha_{j_0}} , \cdots , I_{\alpha_{j_m}}  \} \subset \{ I_{\alpha_0} , \cdots , I_{\alpha_k} \}$. Removing intervals can only increase the minimum left endpoint and/or decrease the maximum right endpoint and therefore $\delta_{\{ I_{\alpha_{j_0}} , \cdots , I_{\alpha_{j_m}}  \}} \le \delta_{\{ I_{\alpha_0} , \cdots , I_{\alpha_k} \}} \le \Lambda$. Furthermore we have
\be
\bigcap_{r=0}^m S_{\alpha_{j_r}} \supseteq \bigcap_{j=0}^k S_{\alpha_j} \neq \emptyset 
\ee
due to the properties of the intersection of sets. Therefore $\mathsf{K}_{\theta,\Lambda}$ is an abstract simplicial complex.

Now that we have built a simplicial complex we can define its homology. We construct the space of $k$-chains by taking formal linear combinations $c = \sum_i a_i \, \sigma_i$, where $\sigma_i$ is a $k$-simplex. We take the coefficients $a_i$ to be in a field $\mathbb{K}$, for example a cyclic group $\zed_p$ with $p$ prime. Given a $k$-simplex $\sigma$, its boundary is determined by its $(k-1)$-subsimplices $\tau \subset \sigma$. To be concrete assume that we fix an ordering of the vertices, so that we can denote a $k$-simplex as $\sigma = [v_0 , \dots , v_k]$. Then we introduce the boundary operator $\partial_k \, : \, C_k \longrightarrow C_{k-1}$ via
\begin{equation}
\partial_k \left( [v_0 , v_1 , \cdots , v_k] \right) = \sum_{i=0}^k (-1)^i [v_0 , \cdots , \hat{v}_i , \cdots , v_k] \, ,
\end{equation}
where the hatted variables are omitted from the sum. One can verify that $\partial_k \partial_{k+1} = 0$, so $\left( C_\bullet , \partial \right)$ is a chain complex. The homology $H_k (C_\bullet ; \mathbb{K})$ is the quotient $Z_k / B_k$ where $Z_k = \ker \, {\partial_k}$ is the space of $k$-cycles and $B_k = \mathrm{im} \, {\partial_{k+1}}$ the space of $k$-boundaries. The Betti numbers are the dimensions of the homology spaces $\beta_k = \dim H_k = \dim Z_k - \dim B_k$ and according to our proposal contain information about the topology of the underlying spacetime; roughly $\beta_k$ measures the number of independent holes of dimension $k$.

We want to interpret the homology of this complex as the homology of the part of spacetime probed by the observer. Let us argue for this interpretation. Consider first a situation where spacetime is fixed and the observer simply performs measurements during a time interval. In this case the algebra of observables is an undeformed type $\mathrm{III}$ algebra and to be precise the whole discussion about measurements of Section \ref{measurements} should be reformulated; let us be heuristic and ignore this issue. Then to a measurement the observer can associate an open set in spacetime, for example a certain subset of the timelike tube envelope associated with the time interval during which the measurement took place, or of the causal diamond; for example the observer can use Hamiltonian evolution to retrace the source of an emitted particle to a certain region of spacetime. To be concrete the observer can consider open sets whose projection on a spacelike surface are balls of a certain radius. In this case if a detector is active in two sufficiently close measurement events the corresponding open sets are intersecting. For example if we are taking open sets in causal diamonds and we are measuring the same physical phenomenon, this will follow by causality for sufficiently near time intervals. In this case our construction is expected to reduce to the \v{C}ech construction in algebraic topology where the simplicial complex is given by the nerve of the covering\footnote{Of course when we speak about a covering we don't mean a covering of the whole manifold, but only of the portion of spacetime accessible to the observer during the measurement process. It is a responsibility of the judicious observer to cover as much of spacetime as possible before drawing conclusions.} (where $(k+1)$-tuple overlaps correspond to $k$-simplices). For this construction to work one needs to ensure that we have a \textit{good covering}. 

Let us illustrate this with one example. Consider an $S^1$ and imagine we have a covering consisting of two open sets at the north and south ``poles''. Then the intersection of the two open sets is not connected and in particular not contractible. In this case the nerve construction fails. One can avoid this problem by considering a finer covering: already choosing three contractible open sets arranged cyclically with only pairwise intersections, each overlap is contractible and the nerve has the correct homology. It is natural to assume that the observer can avoid this problem resorting to a finer covering, by performing more measurements.

Now let us include dynamical gravitons. Spacetime is not fixed anymore but fluctuates due to perturbative quantum gravity corrections. In particular we cannot anymore talk about regions of spacetime without additional information. The concept of a ball of given radius cannot be defined invariantly. On the other hand the measurement protocol which we have described still makes sense, since all the observer needs is the algebra of observables defined along their worldline. We therefore propose that our construction becomes a stand-in for the ordinary nerve construction when spacetime regions become difficult to define. In particular we propose that the homology of the complex constructed from local measurement captures topological data of spacetime (at the macroscopic scale accessible to the observer). The latter is well defined even when geometry fluctuates, as long as fluctuations are not so wide as to cause topology changing transitions. This will not happen in the perturbative regime. Finally what replaces the good cover assumption is the assumption that the set of measurements carried out by the observer is sufficiently fine that in the semiclassical limit its associated regions form a good cover. To be precise, we are not talking about a good cover for all of spacetime, but only for the portion of spacetime accessible to the observer via measurements.

The dependence on the finer details of the measurement process is captured by the functoriality of the construction. Indeed let us assume that $\mathsf{K}_1$ and $\mathsf{K}_2$ are two simplicial complexes obtained as above. For examples such complexes could differ by the times at which measurement events take place, or by the threshold $\Lambda$ or the measure sensitivity $\theta$. Whenever change in these parameters induces a simplicial map $f : \mathsf{K}_1 \longrightarrow \mathsf{K}_2$, this map will induce a chain map $C_\bullet (f) : C_\bullet (\mathsf{K}_1) \longrightarrow C_\bullet (\mathsf{K}_2)$ at the level of chain complexes, leading to the commutative diagram
\begin{equation}
\xymatrix@C=8mm{  
 \cdots  \ar[r]  & C_p (\mathsf{K}_1) \ar[r]^{\partial_p^{\mathsf{K}_1}}  \ar[d]^{C_p (f)} & C_{p-1} (\mathsf{K}_1) \ar[r] \ar[d]^{C_{p-1} (f)} & \cdots \\    
 \cdots  \ar[r]  & C_p (\mathsf{K}_2) \ar[r]^{\partial_p^{\mathsf{K}_2}} & C_{p-1} (\mathsf{K}_2) \ar[r] & \cdots
 } \, .
\end{equation}
This map $f$ induces an homomorphism at the level of the homologies $ f_\star : H_\bullet (\mathsf{K}_1 ; \mathbb{K}) \longrightarrow H_\bullet (\mathsf{K}_2 ; \mathbb{K}) $. For example it is clear that if the map $f$ corresponds only to a small change of the time intervals $(a_\alpha , b_\alpha)$, the combinatorics remains unchanged and the induced homomorphism is actually an isomorphism. More interesting is the case where the map $f$ between $\mathsf{K}_1$ and $\mathsf{K}_2$ is an inclusion, obtained for example by changing the sensitivity threshold $\theta$. We will use this fact in the next Section while discussing persistent homology.

We conclude this discussion with a few comments:
\begin{itemize}
\item The observer can fail! There will be situations where, whether for an uneven distribution of measurement events or an unphysical value of some of the parameters, the observer will fail to detect certain topological information about the surrounding spacetime. There can also be cases in which the instrumental setup is insufficient.
\item We are systematically neglecting the backreaction of the measurement device. This is of course unrealistic and a complex experimental apparatus cannot be modelled with a simple observer Hamiltonian, for example a clock even if interacting with its surroundings. However this is justified if we only care about topological information. While the measurement device will distort the geometry, it will not change the spacetime topology. Therefore for the purpose of constructing the above simplicial complex modelling the observer with a simple Hamiltonian is not an issue. In other words we expect that backreaction can change the details of the complex but not its homology. If however we want to generalize the construction to capture geometric information one would need a more careful approach.
\end{itemize}

\subsection{An illustrative example} Let us go through a simplified example to aid the intuition. 

\begin{wrapfigure}{l}{0.35\textwidth}  
\includegraphics[width=0.35\textwidth]{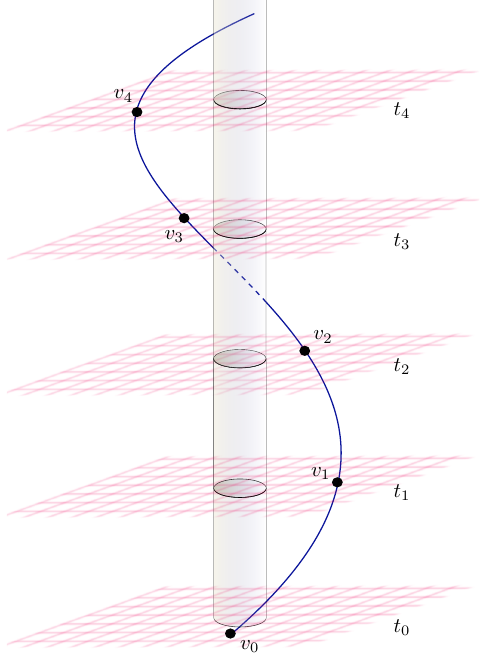}
\caption{The observer moves in spacetime around a ``hole'' and performs five sharp measurements at five different instants}
\label{STpropFig}
\end{wrapfigure}

Let us assume the observer propagates in (2+1)-dimensional spacetime along a timelike curve as in figure \ref{STpropFig}. We illustrate the previous construction in two simplified settings, where the observer circles around a ``hole'' or around a smooth section of spacetime, topologically a disk. To be concrete let us assume that in both cases the observer decides to perform five measurements and that all these measurements at least one detector has an above-threshold response within the temporal scale $\Lambda$. We assume that $\Lambda$ is large enough that the observer has traversed the loop once. To the five events there correspond five vertices $\{ v_0 , v_1 , v_2 , v_3 , v_4 \}$.

We assume the following highly idealized, but illustrative, situation. The observer has a measuring system equipped with four detectors $\cE_{(i)}$ with $i=\textsc{R,B,G,V}$. The observer is measuring focused radiation coming from four distinct sources. To keep things simple, we assume that each source is detected by a distinct detector and is labelled accordingly. In a physical realistic situation we don't expect a simple correspondence between physical effects and detectors. The sources could be stars or other astronomical objects emitting electromagnetic radiation or weakly interacting particles.

We will now compute the relevant simplicial homology in both cases, where the spatial section has a hole or is topologically a disk, working over a field \(\mathbb{K}\). 

\begin{figure}[h]  
\centering
\includegraphics[width=0.5\textwidth]{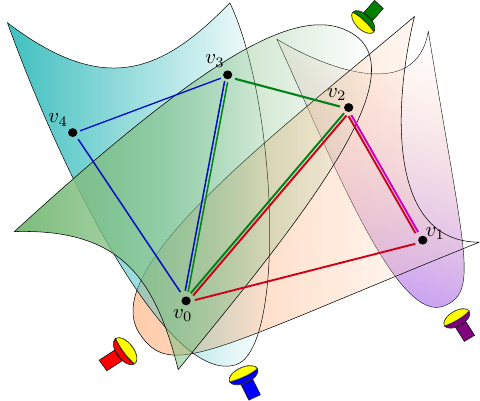}
\caption{The observer records four focused radiation sources (green, blue, red and violet) with five measurements $v_i$, with $i=0,\dots,4$. The timelike dimension is omitted. The red source is detected at measurement $v_0$, $v_1$ and $v_2$, the violet at $v_1$ and $v_2$, the green at $v_0$, $v_2$ and $v_3$, and the blue at $v_0$, $v_3$ and $v_4$. Each measurement is connected by an edge if one or more detectors are active at both events. For illustrative purposes the edges are colored; some of them are doubled for clarity but should really be thought of as the same edge.}
\label{diskFig}
\end{figure}

Consider first the case of the disk. We assume that the detectors are active with the following pattern
\begin{align}
S (v_0) &= \{ {\textsc{R}}  ,    {\textsc{B}} , {\textsc{G}}  \} \, , \qquad
S (v_1) = \{ {\textsc{R}}  ,    {\textsc{V}}   \} \, , \\ \nonumber
S (v_2) &= \{ {\textsc{R}}  ,    {\textsc{G}} ,{\textsc{V}}  \} \, , \qquad
S (v_3) = \{ {\textsc{B}}  ,    {\textsc{G}}  \} \, , \qquad
S (v_4) = \{ {\textsc{B}}   \} \, .
\end{align}
We have introduced the more compact notation $S (v_i)$ to denote the set of labels $S_{\alpha_i}$ associated with the vertex $v_{\alpha_i}$. For example the source \textsc{R} is detected by the measurement events at $v_0$, $v_1$ and $v_2$. A pictorial cartoon of the situation is illustrated in Figure \ref{diskFig}. We omit the timelike direction and project every event to a two dimensional spacelike plane, even if the measurements happen at distinct time intervals.

The shaded areas correspond to the spacetime regions reached by the radiation emitted by the corresponding source. If a measurement event, a vertex, is in one of these regions the corresponding detector is active. When the same detector is active in two distinct measurement event, the corresponding vertices are joined by an edge. Similarly when three events record the same detector active, we have a triangle. Edges are drawn of different colors for sake of clarity. Some edges are doubled to show both colors but this is just an artefact of the presentation and they should be really considered as a single edge.

It is important to stress that Figure \ref{diskFig} is only supposed to be a visual aid about what is going on. In reality spacetime geometry is fluctuating and we cannot really talk about ``regions affected by the radiation'', but only about the measurement events along the observer worldline and the corresponding active detectors. The reader should mind that only the vertices and the simplices are actually defined. Simplices are abstract combinatorial objects and concepts such as ``the length of an edge'' are meaningless.

The simplicial complex is constructed by recording the pattern of measures and their correlations. For example by keeping track of the activity of the detectors $\cE_{(\textsc{R})}$, $\cE_{(\textsc{G})}$ and $\cE_{(\textsc{B})}$ we find the three triangles
\be \label{disktriangles}
d_0 = [v_0v_1v_2],\qquad d_1 = [v_0v_2v_3],\qquad  d_2 = [v_0v_3v_4],
\ee
together with all their faces. The corresponding simplicial complex has therefore five vertices, seven edges and three triangles faces, corresponding to the chain groups
\begin{equation}\label{eq:chain_groups_disk}
C_0(K_{\mathrm{disk}};\mathbb{K})\cong \mathbb{K}^5,\qquad
C_1(K_{\mathrm{disk}};\mathbb{K})\cong \mathbb{K}^7,\qquad
C_2(K_{\mathrm{disk}};\mathbb{K})\cong \mathbb{K}^3.
\end{equation}
We choose the ordered bases
\begin{align}
C_0 &:\ (v_0,v_1,v_2,v_3,v_4), \cr
C_1 &:\ (e_{01},e_{12},e_{23},e_{34},e_{40},e_{02},e_{03}), \cr
C_2 & :\ (d_0,d_1,d_2).
\end{align}
where $e_{ab}$ denotes the oriented edge $[v_a v_b]$. The boundary maps can be written explicitly as
\begin{align}
\label{eq:boundary1}
\partial_1([v_av_b]) & =v_b-v_a,
\\
\label{eq:boundary2}
\partial_2([v_a v_bv_c]) &=[v_av_b]+[v_bv_c]-[v_av_c].
\end{align}
with orientation $a < b < c$. Now by using \eqref{eq:boundary2}, we compute the boundaries of the triangles \eqref{disktriangles}
\begin{equation}\label{eq:triangle_boundaries}
\begin{aligned}
\partial_2(d_0)&=e_{12}-e_{02}+e_{01},\\
\partial_2(d_1)&=e_{23}-e_{03}+e_{02},\\
\partial_2(d_2)&=e_{34}+e_{40}+e_{03}.
\end{aligned}
\end{equation}
Therefore the linear operator $\partial_2$ is represented in this basis by the matrix
\begin{equation}\label{eq:matrix_boundary2_disk}
\partial_2
=
\begin{pmatrix}
 1 & 0 & 0\\
 1 & 0 & 0\\
 0 & 1 & 0\\
 0 & 0 & 1\\
 0 & 0 & 1\\
-1 & 1 & 0\\
 0 &-1 & 1
\end{pmatrix} \, ,
\end{equation}
which we denote with the same symbol by partial abuse of notation. We see immediately that the three columns are linearly independent and therefore $\rank \, \partial_2 = \dim \mathrm{im} \, \partial_2 = 3$.

Similarly from \eqref{eq:boundary1} one finds
\begin{equation}\label{eq:matrix_boundary1_disk}
\partial_1
=
\begin{pmatrix}
-1 & 0 & 0 & 0 & 1 & -1 & -1\\
 1 &-1 & 0 & 0 & 0 &  0 &  0\\
 0 & 1 &-1 & 0 & 0 &  1 &  0\\
 0 & 0 & 1 &-1 & 0 &  0 &  1\\
 0 & 0 & 0 & 1 &-1 &  0 &  0
\end{pmatrix}.
\end{equation}
By row-reduction we can check that this matrix has rank 4. Therefore by the nullity-rank theorem
\begin{equation}\label{eq:ker_dim_disk}
\dim\ker \, \partial_1 =7-4=3.
\end{equation}
Hence
\begin{equation}\label{eq:H1_disk}
\dim H_1(K_{\mathrm{disk}};\mathbb{K})
=
\dim\ker \partial_1-\dim\mathrm{im} \, \partial_2 
=
3-3=0.
\end{equation}
Since the three triangle boundaries are independent, \(\partial_2\) is injective, and therefore
$
H_2(K_{\mathrm{disk}};\mathbb{K})=0
$. Moreover, because the complex is connected,
$
\dim H_0(K_{\mathrm{disk}};\mathbb{K})=1.
$
To summarize the Betti numbers of the disk complex are
\begin{equation}\label{eq:betti_disk}
\beta_0=1,\qquad \beta_1=0,\qquad \beta_2=0 \, ,
\end{equation}
as expected.

We now consider the case where the loop encloses a hole. The presence of the hole obstructs the propagation of signals: certain measurement event will be shielded by the presence of the hole. We should not think of the hole as some physical obstruction to the radiation, for example a planet shielding electromagnetic radiation (this case could be avoided by considering for example neutrino or other weakly interacting particles), but as a genuine spacetime topological feature. This definition also includes black holes since the horizon indeed would not let any entering radiation escape;  in this case the topological feature is really the presence of the singularity in the interior.

\begin{figure}[h] 
\centering
\includegraphics[width=0.5\textwidth]{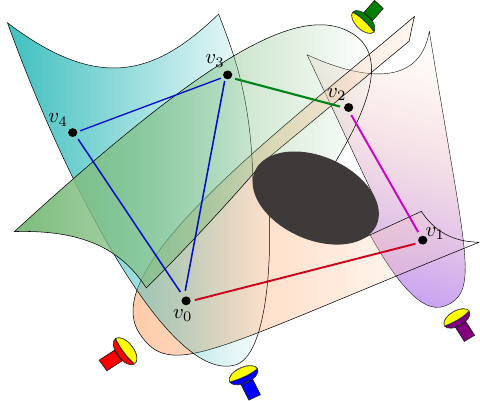}
\caption{Now spacetime is topologically non-trivial. The presence of the hole prevents radiation emitted from the sources to propagate through it. As a consequence the situation is different from before and now no ``red'' radiation is detected at $v_2$ and no ``green'' radiation is detected at $v_0$. From this information the observer is able to reconstruct the presence of the hole.}
\label{holeFig}
\end{figure}

We can for example take the pattern of measures
\begin{align}
S (v_0) &= \{ {\textsc{R}} ,{\textsc{B}}  \} \, , \qquad
S (v_1) = \{ {\textsc{R}}  ,    {\textsc{V}}   \} \cr
S (v_2) &= \{ {\textsc{V}}  ,    {\textsc{G}}  \} \, , \qquad
S (v_3) = \{ {\textsc{G}}  ,    {\textsc{B}}  \} \, , \qquad
S (v_4) = \{ {\textsc{B}}  \} \, .
\end{align}
corresponding to the physical situation illustrated in Figure \ref{holeFig}. Note that the presence of the hole results in certain detectors not being active at certain times due to the position of the observer. The set of recorded events is therefore a subset of the case of the disk.
We will see that this means that the measurement record detects adjacency between subsequent events as the observer moves around the loop, but the hole serves as an obstruction to higher-order overlaps that could fill the interior. 

The complex \(K_{\mathrm{hole}}\) has the same five vertices as before. From the pattern of intersections $S (v_i) \cap S (v_j)$ we see the following edges
\be
[v_0v_1],\quad [v_1v_2],\quad [v_2v_3],\quad [v_0v_3],\quad [v_3v_4],\quad [v_0v_4] \, .
\ee
There is one nonempty triple intersection, $S(v_0)\cap S(v_3)\cap S(v_4)=\{\cE_{(\textsc{B})}\}$,
so the only surviving two-simplex is $d_2 = [v_0v_3v_4]$. All other triple intersections are empty, and similarly no fourfold intersection is nonempty. Therefore the chain groups are
\begin{equation}\label{eq:chain_groups_hole}
C_0(K_{\mathrm{hole}};\mathbb{K})\cong \mathbb{K}^5 \, , \qquad
C_1(K_{\mathrm{hole}};\mathbb{K})\cong \mathbb{K}^6 \, , \qquad
C_2(K_{\mathrm{hole}};\mathbb{K})\cong \mathbb{K} \, .
\end{equation}
Now we chose the ordered bases
\be
C_0:\ (v_0,v_1,v_2,v_3,v_4),\qquad
C_1:\  (e_{01},e_{12},e_{23},e_{03},e_{34},e_{04}) \, .
\ee
The matrix representing the action of \(\partial_1:C_1\to C_0\) is
\begin{equation}\label{eq:matrix_boundary1_hole}
\partial_1
=
\begin{pmatrix}

-1 & 0 & 0 & -1 & 0 & -1\\

 1 &-1 & 0 &  0 & 0 &  0\\

 0 & 1 &-1 &  0 & 0 &  0\\

 0 & 0 & 1 &  1 &-1 &  0\\

 0 & 0 & 0 &  0 & 1 &  1

\end{pmatrix}.
\end{equation}
Proceeding as before we find that $\mathrm{rank} \, \partial_1 = 4$ and therefore by the rank-nullity theorem $\dim\ker \, \partial_1 =6-4=2$. 
The boundary of the unique two-simplex is given by
\begin{equation} \label{boundaryt3}
\partial_2 d_2
=
[v_3v_4]-[v_0v_4]+[v_0v_3]
=
e_{34}-e_{04}+e_{03} \, ,
\end{equation}
which is nonzero. Therefore $\mathrm{rank} \, \partial_2 =1$ and hence $\dim H_1(K_{\mathrm{hole}};\mathbb{K})
=
\dim\ker \partial_1 -\dim\mathrm{im} \, \partial_2 
=
2-1=1$. Finally since
$
H_2(K_{\mathrm{hole}};\mathbb{K})=\ker \, \partial_2=0
$, we find the following Betti numbers
\be
\beta_0=1,\qquad \beta_1=1,\qquad \beta_2=0.
\ee
For example a  generator of the nontrivial first homology class is represented by the cycle
\be
z=e_{01}+e_{12}+e_{23}-e_{03}.
\ee
Indeed,
$
\partial_1 z=0,
$
and \(z\) is not a boundary, since the only two-boundary is \eqref{boundaryt3}. Therefore the complex  detects a single one-dimensional hole via its homology.

In this example we have seen explicitly how from a series of measurements the observer can extract topological information about spacetime. The example was purposely very stylised to illustrate the construction. For example by looking at the pictures, it seems that the observer could miss the presence of the hole if the measurements were distributed differently in time. This issue is of course well known in computational topology \cite{Carlsson:2009zke}, when one addresses similar problems such as image reconstruction, and the solution is to increase the data set; in the language of the previous section this is equivalent to the ``good cover'' condition. This illustrates concretely the possibility that an insufficiently rich set of measurements could fail to reconstruct the correct topology. In a more realistic setup we expect that the observer needs to perform a large number of measurements to detect properly topological features of spacetime. 

\section{Topological persistence} \label{persistence}

The previous construction suffers from certain ambiguities due to the dependence on the parameters $\theta$ and $\Lambda$. Operationally the observer records the measurement data and then chooses the parameters $\theta$ and $\Lambda$. But there might be spurious effects. For example if $\Lambda$ is too big there might be overlaps due to coincidental but unrelated physical effects. Similarly a too high value of $\theta$ might miss physical information. In the examples in the previous Section we assumed these parameters were somehow fixed at some optimal value. A natural way to bypass these problems is to let these parameters vary and work with filtered complexes. This leads to the setup of persistent homology which singles out the topological features which are more robust as $\theta$ or $\Lambda$ vary. The ideas and concepts of persistent homology are nicely exposed in \cite{Carlsson:2009zke}. See also \cite{Cirafici:2015sdg,Cirafici:2015pky,Cole:2018emh} for applications in string theory.

\subsection{Measurements and filtered complexes}

If we now vary the two parameters $\theta$ and $\Lambda$, how does the complex change? Let us begin with $\theta$. To make things clearer we add explicitly the $\theta$ dependence to the notation.

If $\theta_1 \le \theta_2$, then $S_\alpha (\theta_2) \subseteq S_\alpha (\theta_1)$ since increasing the sensitivity threshold at most removes active detectors. This means that for any finite set $\{ \alpha_0 , \dots , \alpha_k \}$ we have 
\be
\bigcap_j S_{\alpha_j} (\theta_2)  \subseteq \bigcap_j S_{\alpha_j } (\theta_1) 
\ee
or equivalently
\be
\bigcap_j S_{\alpha_j} (\theta_2) \neq\varnothing
\;\Longrightarrow\;
\bigcap_j S_{\alpha_j}(\theta_1)\neq\varnothing.
\ee
so that $\mathsf{K}_{\theta_2 , \Lambda} \subseteq \mathsf{K}_{\theta_1 , \Lambda}$. The complex grows when $\theta$ decreases. In particular these inclusions induce a natural filtration. If $\Lambda$ is fixed and we take a decreasing sequence $\theta_1 \ge \theta_2 \ge \cdots \ge \theta_m$, we have 
\be \label{thetafilt}
\mathsf{K}_{\theta_1 , \Lambda} \subseteq \mathsf{K}_{\theta_2 , \Lambda} \subseteq \cdots \subseteq \mathsf{K}_{\theta_m , \Lambda}
\ee

Imagine now varying $\Lambda$, with $\theta$ fixed. By definition, if $\Lambda_1 \le \Lambda_2$, then $\delta_{ \{ I_{\alpha_0} , \dots , I_{\alpha_k} \}} \le \Lambda_1$ implies $\delta_{ \{I_{\alpha_0} , \dots , I_{\alpha_k} \}} \le \Lambda_2$ so that $\mathsf{K}_{\theta , \Lambda_1} \subseteq \mathsf{K}_{\theta , \Lambda_2}$. The complex now grows as we increase $\Lambda$. If we fix $\theta$ and take an increasing sequence $\Lambda_1 \le \Lambda_2 \le \cdots \le \Lambda_m$, now the filtration reads
\be \label{Lambdafilt}
\mathsf{K}_{\theta , \Lambda_1} \subseteq \mathsf{K}_{\theta , \Lambda_2} \subseteq \cdots \subseteq \mathsf{K}_{\theta , \Lambda_m} \, ,
\ee
assuming that $\Lambda_1$ is larger than the maximal duration of a measurement event.

\subsection{Persistent homology}

The topological information contained in a filtered complex is captured by its persistent homology. To be concrete, let us consider the filtration \eqref{Lambdafilt}. Taking the $i$-th homology gives
\begin{equation} \label{Npersistence}
H_i (\mathsf{K}_{\theta , \Lambda_1} ; \mathbb{K})  \longrightarrow H_i (\mathsf{K}_{\theta , \Lambda_2}  ;  \mathbb{K}) \longrightarrow H_i (\mathsf{K}_{\theta , \Lambda_3} ;  \mathbb{K}) \longrightarrow \cdots \, ,
\end{equation}
where the maps are the homomorphisms induced in homology by the natural inclusions of the original filtration. This is an example of a persistence module.

Functoriality of the construction guarantees that we can follow a homology class as it is generated for certain values $\Lambda_a$ and it disappears (becoming trivial) at a subsequent value $\Lambda_b$. The intervals $[\Lambda_a,\Lambda_b )$ characterize the homology class and determine a so-called \textit{barcode}.

A more precise characterization of a persistence module is as a collection of finite-dimensional vector spaces $\{ V_i \}_i$ over a field $\mathbb{K}$ and homomorphisms $\rho_{i,j} \, : \, V_i \longrightarrow V_j$ with $i \le j$, such that $\rho_{k,j} \circ \rho_{i,k} = \rho_{i,j}$ whenever $i \le k \le j$, and $\rho_{i,i} = \mathrm{id}_{V_i}$. Persistence modules can be decomposed as
\begin{equation}
\{ V_i \}_i \simeq \bigoplus_{j=1}^N \IK \, (m_j , n_j) \, ,
\end{equation}
for some $N$. Here $(m_j , n_j)$ are a collection of non-negative integers with  $0 \le m_j \le n_j$ and
$\IK \, (m , n)$ denotes the persistence module defined by
\begin{equation}
\IK \, (m,n)_i = 
\begin{cases}
& \IK \qquad m \le i < n \\
& 0, \qquad $otherwise$ \\
\end{cases} \, .
\end{equation}
In other words $\IK \, (m_j , n_j)$ is an interval module, which assigns the vector space $\IK$ to the interval $[m_j , n_j)$, so that the morphism $\rho_{i,j}$ is just the identity for $m \le i \le j < n$ (allowing $n = \infty$). The multiset of intervals $[m_j , n_j)$ is the barcode, represented graphically as a collection of $N$ bars; each bar corresponds to a homology class which is created at $m_i$ and disappears at $n_i$.

The precise interpretation of persistent homology depends on the filtration at hand. In the context of the $\Lambda$ filtration we interpret homology classes which persist, in the sense of being long-lived, for small values of $\Lambda$ to correspond to truly topological features of spacetime; features that appear only after $\Lambda$ has become very large are likely to be the result of accidental correlations caused by reusing the same detectors at widely separated instants. In other words if we wait long enough, the same detector can be active at two different instants even if we are not measuring the same physical phenomenon. Persistence is meaningful only while $\Lambda$ is compatible with the causal structure of propagating signals and features outside this range should be discarded. In this sense there is a value $\Lambda_{\mathrm{max}}$ above which the filtration is no longer admissible and classes should be considered ``long-lived'' only within this interval. In the case of the $\theta$ filtration, classes which only appear for low values of the threshold can be interpreted as noise, while long lived classes correspond to robust features. However values of $\theta$ close to one are most likely to miss relevant topological features.

The central idea is therefore not to pick a specific value of the $\theta$ and $\Lambda$ parameters, but to let them vary. The observer will then determine which homology classes correspond to topological features of spacetime and which are most likely just artefacts of the measurement protocol. Topological features that persist under further refinements are natural candidates for information which is insensitive to the measurement details.

\subsection{The example revisited}

Let us revisit our previous example from the point of view of the $\Lambda$ filtration. As before we have five measurement events corresponding to the vertex set $\{  v_0, v_1, v_2, v_3, v_4 \}$. Let us assume that these events are equally spaced in time and happen at proper times $t_j = t_0 + j \, \tau$ with $j=0,\dots,4$ and $\tau > 0$. Since we assume that the measurements are instantaneous we shall use the simplified notation $\delta (\{v_i , \dots , v_j\})$ to indicate the interval of time lapsed between the measurement $v_i$ and $v_j$. We fix the threshold $\theta$ which determines when a detector $S (v_i)$ is active and let vary the parameter $\Lambda$, which determines when the observer accepts that two events are correlated. As before the family \(\{K_\Lambda\}_{\Lambda\ge0}\) is a filtration.

For simplicity let us consider only the case where a spacelike section of the $(2+1)$-dimensional spacetime contains a hole (the disk case is very similar). For values of the parameter $\Lambda \in [0 , \tau)$ no two distinct measurements are close enough in time to form an edge and therefore $\mathsf{K}_\Lambda$ only consists of the five vertices. Therefore $H_0(K_\Lambda;\mathbb{K})\cong \mathbb{K}^5$ and $H_1(K_\Lambda;\mathbb{K})=0$.

At the next step for $\Lambda \in [\tau , 3 \tau)$ the active detectors are
\begin{align}
S(v_0)\cap S(v_1)&=\{{\textsc{R}}\}, \qquad
S(v_1)\cap S(v_2)=\{{\textsc{V}}\}, \cr
S(v_2)\cap S(v_3)&=\{{\textsc{G}}\}, \qquad
S(v_3)\cap S(v_4)=\{{\textsc{B}}\}
\end{align}
and the four consecutive edges $[v_0v_1]$, $[v_1v_2]$, $[v_2v_3]$ and $[v_3v_4]$ appear and form a connected tree, so that now $H_0(K_\Lambda;\mathbb{K})\cong \mathbb{K}$ and $H_1(K_\Lambda;\mathbb{K})=0$.

When we reach $\Lambda = 3 \tau$ a new edge appears since $S(v_0)\cap S(v_3)=\{{\textsc{B}}\}$ and now $\delta(\{v_0,v_3\})=t_3-t_0=3\tau$. Now the edge $[v_0 v_3]$ enters the complex and the one-cycle
\be
z=[v_0v_1]+[v_1v_2]+[v_2v_3]-[v_0v_3]
\ee
appears. We see immediately that 
\be
\partial_1 z
=
(v_1-v_0)+(v_2-v_1)+(v_3-v_2)-(v_3-v_0)=0
\ee
and therefore $z \in \ker \partial_1$. Since no two-simplices are present, $\mathrm{im} \, \partial_2 = 0$ and therefore $z$ defines a nontrivial homology class. Therefore we now have $H_0(K_\Lambda;\mathbb{K})\cong \mathbb{K}$ and
$H_1(K_\Lambda;\mathbb{K})\cong \mathbb{K}$. 


The situation remain unchanged until we reach $\Lambda = 4 \tau$ where two new simplices appear. One is the edge $[v_0v_4]$, since $S(v_0)\cap S(v_4)=\{{\textsc{B}} \}$ and $\delta(\{v_0,v_4\})=t_4-t_0=4\tau$. The other one is the two-simplex $[v_0 v_3 v_4]$ since $S(v_0)\cap S(v_3)\cap S(v_4)
=
\{{\textsc{B}}\}$ and again $\delta(\{v_0,v_3,v_4\})=t_4-t_0=4\tau$. By using the new edge one can form the one cycle $[v_0v_3]+[v_3v_4]-[v_0v_4]$. However this cycle is trivial in homology since it is a boundary: $\partial_2[v_0v_3v_4]
=
[v_3v_4]-[v_0v_4]+[v_0v_3]$.

It remains to be checked that the earlier cycle $z$ is not killed. Since the only two-simplex present is $[v_0 v_3 v_4]$, the vector space $\mathrm{im} \, \partial_2$ is one dimensional and generated by $[v_3v_4]-[v_0v_4]+[v_0v_3]$. However $z$ cannot be a scalar multiple of this boundary since it contains the edges $[v_0 v_1]$, $[v_1 v_2]$ and $[v_2 v_3]$ which are not in $\mathrm{im} \, \partial_2$. Therefore $z \notin \mathrm{im} \, \partial_2$ and its homology class survives for all $\Lambda \ge 4 \tau$.

\begin{wrapfigure}{l}{0.35\textwidth}  
\centering
\includegraphics[width=0.35\textwidth]{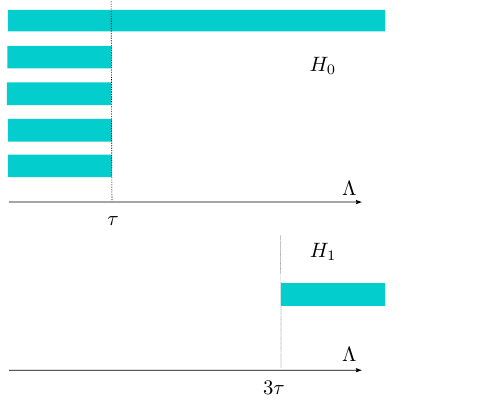}
\caption{An illustration of the barcodes obtained in the example}
\label{barcodeH}
\end{wrapfigure}

At this stage the chain groups are $C_0(K_\Lambda;\mathbb{K})\cong \mathbb{K}^5$, generated by the five vertices, $C_1(K_\Lambda;\mathbb{K})\cong \mathbb{K}^6$, generated by the edges $[v_0v_1]$, $[v_1v_2]$,  $[v_2v_3]$, $[v_0v_3]$, $[v_3v_4]$ and $[v_0v_4]$, and $C_2(K_\Lambda;\mathbb{K})\cong \mathbb{K}$ generated by $[v_0 v_3 v_4]$. It is now easy to compute the homology. Since $\mathrm{rank} \, \partial_1 = 4$, we have that $\dim\ker\partial_1
=
\dim C_1-\mathrm{rank}\, \partial_1
=
6-4=2$. But we have already seen that $\dim \mathrm{im} \partial_2 = 1$ and therefore $\dim H_1(K_\Lambda;\mathbb{K}) = 1$. Finally since the complex contains only one two-simplex, $[v_0 v_3 v_4]$ with nonzero boundary, $H_2(K_\Lambda;\mathbb{K}) = 0$.

The hole produces a one-dimensional class which is born when $[v_0 v_3]$ appears at a scale $3 \tau$ and persists as we vary $\Lambda$. In particular it is not killed by the appearance of the triangle $[v_0 v_3 v_4]$. The observer therefore can decide to interpret it as a genuine feature of spacetime. The barcodes computed in this example are illustrated in Figure \ref{barcodeH}.

The computation for the case where spacetime is locally a disk is completely identical for $H_0$ while no nontrivial class appears in $H_1$, and we will not include it.

\section{Some final comments}

In this note we have outlined a prescription which an observer can use to measure the topology of the surrounding spacetime, even when this is fluctuating. The basic idea is to rephrase topological information into a complex constructed out of local measurements of the algebra of observables accessible to the observer along their worldline.

A generic observer has associated with their wordline a background-independent algebra of observables. Upon choosing a background this algebra admits a Hilbert space completion as a von Neumann algebra. The algebra includes the metric, quantized in perturbation theory. As a consequence spacetime is subjected to perturbative quantum fluctuations and geometrical concepts such as lengths or distances become hard to define. Nevertheless the observer can compute the homology of the accessible region of spacetime by using the von Neumann algebra of observables available along their worldline.

Such an algebra is constructed by imposing the Hamiltonian constraint, with respect to the Hamiltonian of the bulk and of the observer. This construction can be generalized to include localized interactions between the observer and the bulk, working order by order in perturbation theory. Under reasonable assumptions the resulting algebra is a type $\mathrm{II}$ von Neumann algebra endowed with a thermal state of maximal entropy. The general properties of this algebra guarantee that the observer has at their disposal enough projections to set up local measurements.

Then the observer can construct a simplicial complex where the vertices are associated with time-localized measurements and the structure of higher-simplices follows from correlations in time between the measurements. We have proposed that such a complex captures topological information about spacetime via its homology. While the construction is highly idealized, the main ideas can form the core of a more realistic measurement process, where the observer can incorporate more realistic effects. We have argued that by using techniques from persistent homology the observer can potentially remove certain ambiguities in the construction.

There are a few generalizations that we believe are worth investigating. First of all it would be very interesting if one could actually prove rigorously that our construction computes the topology of the spacetime accessible to the observer.  A fascinating problem to investigate is to understand if a version of this construction could be carried on at the level of the background-independent algebra. Perhaps a more refined version could be even used to extract or define geometric information about a fluctuating quantum mechanical spacetime.

The correlations between measurements that give rise to the simplicial complex are essentially classical. It would be interesting to see if the construction can be generalized using actual correlation functions or $n$-time measurements (as in \cite{Cirafici:2024ccs}). The obvious problem is that these would not be symmetric under generic permutations of their entries and therefore won't define a simplex.

Finally while we have focussed on an observer moving in closed universes, it is natural to wonder if a similar construction can be carried out in asymptotically AdS spaces from boundary observables.

\section*{Acknowledgements} I am supported by \textsc{INFN} via the \textsc{GAST} initiative, and I am a member of \textsc{INDAM} and \textsc{IGAP}. 

\appendix

\section{The Hamiltonian constraint at higher orders} \label{appendix}

In the main text we computed the first-order correction to the relational observable associated with the deformed constraint
\be
\widehat H_\lambda=\widehat H_0+\lambda V.
\ee
For completeness, we now generalize this construction to all orders in \(\lambda\).

We write the full field as an expansion $
\Phi_s 
=
\sum_{n=0}^{\infty}\lambda^n\Phi_s^{(n)},
$
where
$
\Phi_s^{(0)}=\phi (p+s),
$
satisfies
$
[\widehat H_0,\Phi_s^{(0)}]=0.
$
We now want to impose the constraint
$
[\widehat H_\lambda,\Phi_s ]=0
$
at all orders.

Substituting the perturbative expansion and collecting powers of \(\lambda\), one finds
\be
[\widehat H_0,\Phi_s^{(n)}]
=
-[V,\Phi_s^{(n-1)}],
\qquad n\geq 1.
\ee
To solve these equations it is convenient to introduce the interaction-picture operator
\be
V(\tau)
=
e^{i\widehat H_0\tau}
V
e^{-i\widehat H_0\tau}.
\ee
The first-order correction derived in the main text can be written in the compact form
\be
\Phi_s^{(1)}
=
-i
\int_{-\infty}^{0}
d\tau_1\,
[V(\tau_1),\Phi_s^{(0)}] \, ,
\ee
where we have suppressed, and will henceforth always suppress, the damping factor $\e^{\varepsilon \tau_1}$. Iterating the construction, the second-order contribution is
\be
\Phi_s^{(2)}
=
(-i)^2
\int_{-\infty}^{0}
d\tau_1
\int_{-\infty}^{\tau_1}
d\tau_2\,
[V(\tau_1),
[V(\tau_2),\Phi_s^{(0)}]].
\ee
Proceeding recursively, one can obtain the expression for $\Phi_s^{(n)}$ order by order. The full perturbative series is then given by
\be
\Phi_s 
=
\phi (p+s) 
+
\sum_{n=1}^{\infty}
(-i\lambda)^n
\int_{-\infty}^{0}
d\tau_1
\int_{-\infty}^{\tau_1}
d\tau_2
\cdots
\int_{-\infty}^{\tau_{n-1}}
d\tau_n\,
[V(\tau_1),
[V(\tau_2),
\cdots
[V(\tau_n),\phi_s]
\cdots ]] \, ,
\ee
where again all integrals are understood formally. This series has the form of a Dyson expansion. Let us introduce the adjoint action of the interaction
\be
\operatorname{ad}_{V(\tau)}(A)
=
[V(\tau),A] \, .
\ee
Then the perturbative series can be resummed formally as
\be
\Phi_s
=
\mathcal T
\exp
\left(
-i\lambda
\int_{-\infty}^{0}
d\tau\,
\operatorname{ad}_{V(\tau)}
\right)
\phi (p+s) ,
\ee
where time ordering acts on the derivation $\operatorname{ad}_{V(\tau)}$. This provides the formal all-order deformation of the relational observable associated with the perturbed constraint \(\widehat H_\lambda\).

\end{document}